\documentclass[twocolumn,final,epj]{svjour}
\usepackage[utf8]{inputenc}
\usepackage[T1]{fontenc}
\usepackage{graphicx}  
\usepackage{amssymb}
\usepackage{amsmath}
\usepackage{hyperref}
\hypersetup{
    colorlinks=true,       % false: boxed links; true: colored links
    linkcolor=blue,          % color of internal links (change box color with linkbordercolor)
    citecolor=blue,        % color of links to bibliography
    filecolor=magenta,      % color of file links
    urlcolor=cyan           % color of external links
}
\usepackage{color}

\begin{document}

\title{Thermoelectricity of cold ions in optical lattices}

\author{
Oleg V. Zhirov\inst{1,2,3} 
\and
Jos\'e Lages\inst{4}
\and
Dima L. Shepelyansky\inst{5}}

\institute{
Budker Institute of Nuclear Physics,
630090 Novosibirsk, Russia
\and
Novosibirsk State University, 630090 Novosibirsk, Russia
\and
Novosibirsk State Technical University, Novosibirsk, Russia
\and
Institut UTINAM, OSU THETA, CNRS, 
Universit\'e de Bourgogne Franche-Comt\'e, Besançon, France
\and
Laboratoire de Physique Th\'eorique, IRSAMC, 
Universit\'e de Toulouse, CNRS, UPS, 31062 Toulouse, France
}

\titlerunning{Thermoelectricity of cold ions}
\authorrunning{O.V.~Zhirov, J.~Lages and D.L.~Shepelyansky}

%\date{\today}
\date{Dated: 28 January 2019}

\abstract{We study analytically and numerically
the thermoelectric properties of cold ions
placed in an optical lattice. Our results show that
the transition from sliding to pinned phase
takes place at a certain critical amplitude of lattice potential 
being similar to the Aubry transition for the Frenkel-Kontorova model.
We show that this critical amplitude is proportional
to the  cube of ion density that allows to perform
experimental realization of this system at moderate
lattice amplitudes. We show that the Aubry phase
is characterized by the dimensionless Seebeck coefficient 
about 50 and the figure of merit being around 8.
We propose possible experimental investigations of
such system with cold ions and argue that the experiments with
electrons on liquid helium surface can also
help to understand its unusual properties.
The obtained results represent also a challenge for modern
methods of quantum chemistry and material science.
}

%% PACS to be updated or removed
%\PACS{
%{89.75.Fb}{
%Structures and organization in complex systems}
%\and
%{89.75.Hc}{
%Networks and genealogical trees}
%\and
%{89.20.Hh}{
%World Wide Web, Internet}
%}

\maketitle

\section{Introduction}
\label{sec:1}

The Wigner crystal \cite{wigner}  
has been realized with a variety of 
physical systems including
cold ions in radio-frequency traps \cite{walther,dubinrmp},
electrons on a surface of liquid helium \cite{konobook},
quantum wires in solid state systems 
(see e.g. review \cite{matveev}),
and even dusty plasma in space \cite{fortov}.

The Cirac-Zoller proposal to perform quantum computations
with trapped cold ions \cite{zoller} pushed forward
the cold ion investigations with generation of quantum entanglement,
realization of main quantum gates and simple algorithms
(see e.g. review \cite{blatt2008}). In addition
the quantum simulations of various physical 
systems with cold atoms became an independent and important
research direction (see e.g. reviews \cite{wunderlich,blatt2012}).
At present the experiments with a chain of up to 100 cold ions
have been reported \cite{monroe}.

The proposal to study the properties of Wigner crystal 
\cite{wigner} in a periodic optical lattice
with cold trapped ions in one-dimension (1D)
has been introduced in \cite{fki2007}.
The analytical and numerical studies performed there
established the emergence of transition from sliding phase
at weak potential amplitudes to the pinned crystal phase at 
high amplitudes. It was shown that this transition
is the Aubry type transition \cite{aubry}
appearing in the Frenkel-Kontorova model
of a chain of particles connected by springs and 
placed on a periodic substrate \cite{obraun}.
In fact, the properties of ionic chain can be locally described 
by the Chirikov standard map \cite{chirikov}
which properties are directly linked with those
of the Frenkel-Kontorova model.
Thus the sliding phase corresponds to the integrable
Kolmogorov-Arnold-Moser (KAM) curves 
while the pinned Aubry phase corresponds to the cantori
invariant set embedded in the phase space component 
with dynamical chaos. The mathematical properties of such
symplectic maps had been studied in great detail
in the field of dynamical chaos 
(see \cite{aubry,lichtenberg,meiss})
and a variety of their physical application
are highlighted in \cite{chirikov,stmapscholar}.

The proposal \cite{fki2007}
attracted the interest of cold ions experimental groups
\cite{haffner2011,vuletic2015sci}
with the first signs of observation of the Aubry
transition reported by Vuletic group \cite{vuletic2016natmat}
with 5 cold ions in a periodic potential
followed by experiments with a larger number of ions
placed in two chains with an effective
periodic potential created by one chain acting on  another one 
\cite{ions2017natcom}.

The physical properties of the Wigner crystal
in a periodic potential are highly nontrivial and interesting.
It was shown that the quantum model in the pinned phase
has an exponentially large
number of exponentially quasi-degenerate configurations
with instanton quantum tunneling between these configurations
in the vicinity of the vacuum state \cite{fki2007}.
Thus this system represents an example of dynamical
spin-glass model where the exponential
quasi-degeneracy emerges not due to 
external disorder but due to nonlinearity 
and chaos of the underlying dynamical map.
However, in addition to this interesting physics
it has been argued \cite{tosatti1,tosatti2} that this model
captures the main mechanisms of friction 
at nanoscale so that the cold ion experiments
can  represent the microscopic friction emulators
allowing to understand study tribology at nanoscale \cite{tosatti3}.
Also it was shown recently that in the case of an asymmetric
potential there is emergence of the Wigner crystal diode 
current in one and two dimensions \cite{diode}.

Of course, the applications of physics of Wigner crystal 
in a periodic potential to nanofriction is very useful and important 
research direction. But in addition it has been shown that
this system possesses exceptional thermoelectric properties \cite{ztzs}.
The fundamental aspects of thermoelectricity had been 
 analyzed in far 1957 by Ioffe \cite{ioffe1,ioffe2}.
At present the needs of efficient energy usage stimulated extensive 
investigations of various materials with high
characteristics of thermoelectricity as reviewed in
\cite{sci2004,thermobook,baowenli,phystod,ztsci2017}.

The thermoelectricity is characterized by 
the Seebeck coefficient $S=-\Delta V /\Delta T$
(or thermopower) which is expressed through a voltage difference $\Delta V$
compensated by a temperature difference $\Delta T$.
We use units with a charge $e=1$ and the Boltzmann constant $k=1$ 
so that $S$ is dimensionless ($S=1$ corresponds to
$S \approx 88 \rm\mu V/K$ (microvolt per Kelvin)).
The thermoelectric materials are ranked by a figure of merit 
$ZT=S^2\sigma T/ \kappa$ \cite{ioffe1,ioffe2}, 
where $\sigma$ is the electric conductivity,
$T$ is material temperature and $\kappa$ is the thermal conductivity.
For an efficient usage of thermoelectricity 
one needs to find materials
with $ZT > 3$ \cite{sci2004,phystod}. At present the highest $ZT$ 
value observed in material science experiments is $ZT \approx 2.6$ \cite{ztsci2017}.
It has been argued that the materials with an effective reduced dimensionality
favor the high thermoelectric performance \cite{dresselhaus}.
The results obtained in \cite{ztzs} showed that in the Aubry pinned phase
it is possible to reach $ZT > 3$ and $S \gg 1$ while the KAM sliding phase
has low values of  $ZT$ and $S$.
However, in \cite{ztzs} only a case of relatively high
charge density has been considered
which requires high power lasers
for a generation of high amplitude of
periodic potential. In this work
we consider the case of low ion density
showing that in this case only a moderate
potential amplitude is required to reach
$ZT > 3$. Thus we expect that the experiments
with cold ions in optical lattices 
can be used as emulators of thermoelectricity
on nanoscale allowing to understand
the physical mechanisms of efficient
thermoelectricity. 

In our opinion, the deep understanding of these
mechanisms is required to select material with 
high thermoelectric properties.
At present there are a lot of quantum physico-chemistry
numerical computations of thermoelectric parameters
for a variety of real materials (see e.g. \cite{kozinsky,melendez}).
With the advanced computational methods
the band structures of electronic transport are
determined taking into account all atom and electron interactions.
However, after that, the conductivity is computed as for
noninteracting electrons. In contrast,
we argue that the interactions of charges (electrons or ions)
is crucial for their thermoelectric properties of transport
in a periodic potential of atomic structures.
Due to these reasons we believe that the 
theoretical and experimental investigations of
Wigner crystal transport in a periodic potential
is crucial for the understanding of thermoelectricity
at atomic and nanoscales.

We note that in addition to the cold ions experiments
in optical lattices there are also other 
physical systems which can be used as a test bed 
for thermoelectricity at nanoscale.
We consider as rather promising the electrons
on liquid helium surfaces within narrow 
quasi-1D channels \cite{kono1d,konstantinov}
and  colloidal monolayers where the signatures of Aubry
transition has been observed recently \cite{bechingerprx}.

The paper is composed as follows: descriptions of model and numerical methods
are given in Section 2,
the dependence of the Aubry transition on charge density is determined
in Section 3 by the numerical simulations and the reduction 
to the symplectic map and its local analysis via the Chirikov standard map,
the formfactor of the ion structure in a periodic potential
is considered in Section 4,  the Seebeck coefficient $S$ is determined in the KAM sliding
phase and the Aubry pinned phase in Section 5, the dependence of figure of merit $ZT$
on system paremeters is established in Section 6 and the discussion
of the results is given in Section 7.

\section{Model description}
\label{sec:2}

The Hamiltonian of the chain of ion charges in a periodic potential
is
\begin{eqnarray}
\nonumber
H &=& {\sum_{i=1}^N} \left( \frac{{P_i}^2}{2} + V(x_i) \right) + U_C \; ,\\
\nonumber
 U_C &=& \sum_{i > j} \frac{ 1} {\mid x_i - x_j \mid} \; ,\\
V(x_i) &=& - K  \cos x_i  \; .
\label{eq:ham1d}
\end{eqnarray}
Here $x_i,P_i$ are conjugated coordinate and momentum of
particle $i$, and $V(x)$  is an external 
periodic potential of amplitude $K$. The Hamiltonian
is written in dimensionless units
where the lattice period is $\ell=2\pi$
and ion mass and charge are $m=e=1$.
In these atomic-type units 
the physical system parameters are measured in
units:   $r_a= \ell/2\pi$ for length,
$\epsilon_a = e^2/r_a = 2\pi e^2/\ell$ for energy,
$E_{adc} = \epsilon_a/e r_a$  
for applied static electric field,
$v_a=\sqrt{\epsilon_a/m}$ for particle velocity $v$,
$t_a =  e r_a \sqrt{m/\epsilon_a}$ for time $t$.

\begin{figure}[t]
\begin{center}
\includegraphics[width=0.48\textwidth]{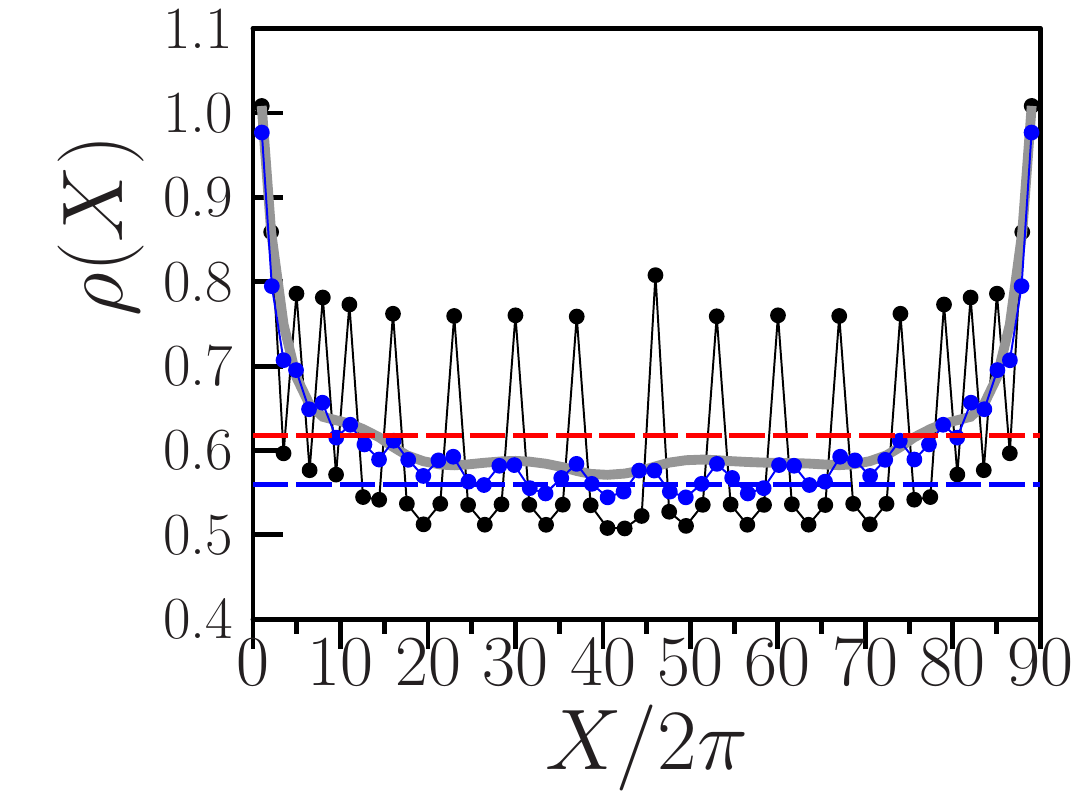}
\end{center}
\caption{\label{fig1} Local ion density distribution is shown along 
the ion chain for
the sliding phase at $K=0.001$ (blue points connected by lines)
and for the pinned phase at $K=0.01$  
(black points connected by lines; the gray curve shows the smoothed distribution).
Here the local density is defined as
$\rho(X) = 2\pi/(x_{i+1}- x_i)$ and $X=(x_{i+1}+x_i)/2$.
The red dashed line corresponds to 
the fraction of golden mean density $\rho(X) = \nu_g -1 = (\sqrt{5}-1)/2 = 0.618...$,
and the blue dashed line corresponds to 
$\rho(X) = \nu =0.56$ obtained from the formfactor data 
(see Fig.~\ref{fig6} below). Here the ratio of $N$ and $ L$ is 
$N/L=55/89 \approx \nu_g \approx 0.618...\;$.
 }
\end{figure}

We note that in this work we consider only the problem of classical
charges. Indeed, as shown in \cite{fki2007} the dimensionless Planck
constant of the system is 
$\hbar_{\rm eff}=$ $\hbar/$ $(e \sqrt{m \ell/2\pi})$.
For a typical lattice period $\ell \approx 1 \rm\mu m  $, $\nu \sim 1$ 
and  ${^{40}}\rm Ca^{+}$ we have $\hbar_{\rm eff} \approx  10^{-5}$.
For electrons on a periodic potential of liquid helium
with the same period $\ell \approx 1 \rm\mu m$
we have $\hbar_{\rm eff} \approx 2 \times 10^{-3}$. Due to this reason
we consider below only the classical dynamics of charges.

Following \cite{ztzs} we model the dynamics of ions in the frame of
Langevin approach (see e.g. \cite{politi})
with the equation of motion being
\begin{equation}
\dot{P}_i = \dot{v}_i= -\partial H/\partial x_i +E_{dc} -\eta P_i+g \xi_i(t)
\; , \;\; \dot{x_i} = P_i  = v_i\; .
\label{eq:langevin}
\end{equation}
The parameter $\eta$  describes phenomenologically the 
dissipative relaxation processes, and 
the amplitude of Langevin force $g$ is given 
by the fluctuation-dissipation theorem $g=\sqrt{2\eta T}$.
We also use particle velocities $v_i=P_i$ (since mass is unity).
The normally distributed random variables $\xi_i$ are 
defined by correlators
$\langle\langle\xi_i(t)\rangle\rangle=0$,
$\langle\langle\xi_i(t) \xi_j(t')\rangle\rangle=\delta_{i j}\delta(t-t')$.
The amplitude of the static force is given by $E_{dc}$.
The equations (\ref{eq:langevin}) are solved numerically
by the 4th order Runge-Kutta integration with a time step $\Delta t$, at
each such a step the Langevin contribution is taken into
account, As in \cite{ztzs} we usually use $\Delta t = 0.02$
and $\eta=0.02$ with the results being not sensitive to these parameters.

The length of the system in 1D case is taken to
be $2\pi L$ in $x$-direction 
with $L$ being the integer number of periods
with periodic 
or hard-wall boundary conditions
for $N$ ions inside the system. 
The dimensionless charge density is $\nu =N/L$ 
being related to the widing number of the KAM curves
in the related symplectic map description 
of equilibrium positions of ions.

\section{Density dependence of the Aubry transition}
\label{sec:3}

The equilibrium static positions of ions in a periodic potential
are determined by the conditions $\partial H/ \partial x_i =0$, $P_i=0$ \cite{aubry,fki2007}.
In the approximation of nearest  neighbor
interacting ions, this leads to the symplectic map for recurrent
ion positions $x_i$
\begin{eqnarray}
p_{i+1} = p_i + K g(x_i) \; , \; \; x_{i+1} = x_i+1/\sqrt{p_{i+1}} \; ,
\label{eq:map}
\end{eqnarray}
where the effective momentum conjugated to $x_i$ is
$p_i = 1/(x_{i}-x_{i-1})^2$ and the kick function
$K g(x_i)= \left.-dV/dx\right|_{x=x_i}$ $ = - K \sin x_i$.

\begin{figure}[t]
\begin{center}
\includegraphics[width=0.48\textwidth]{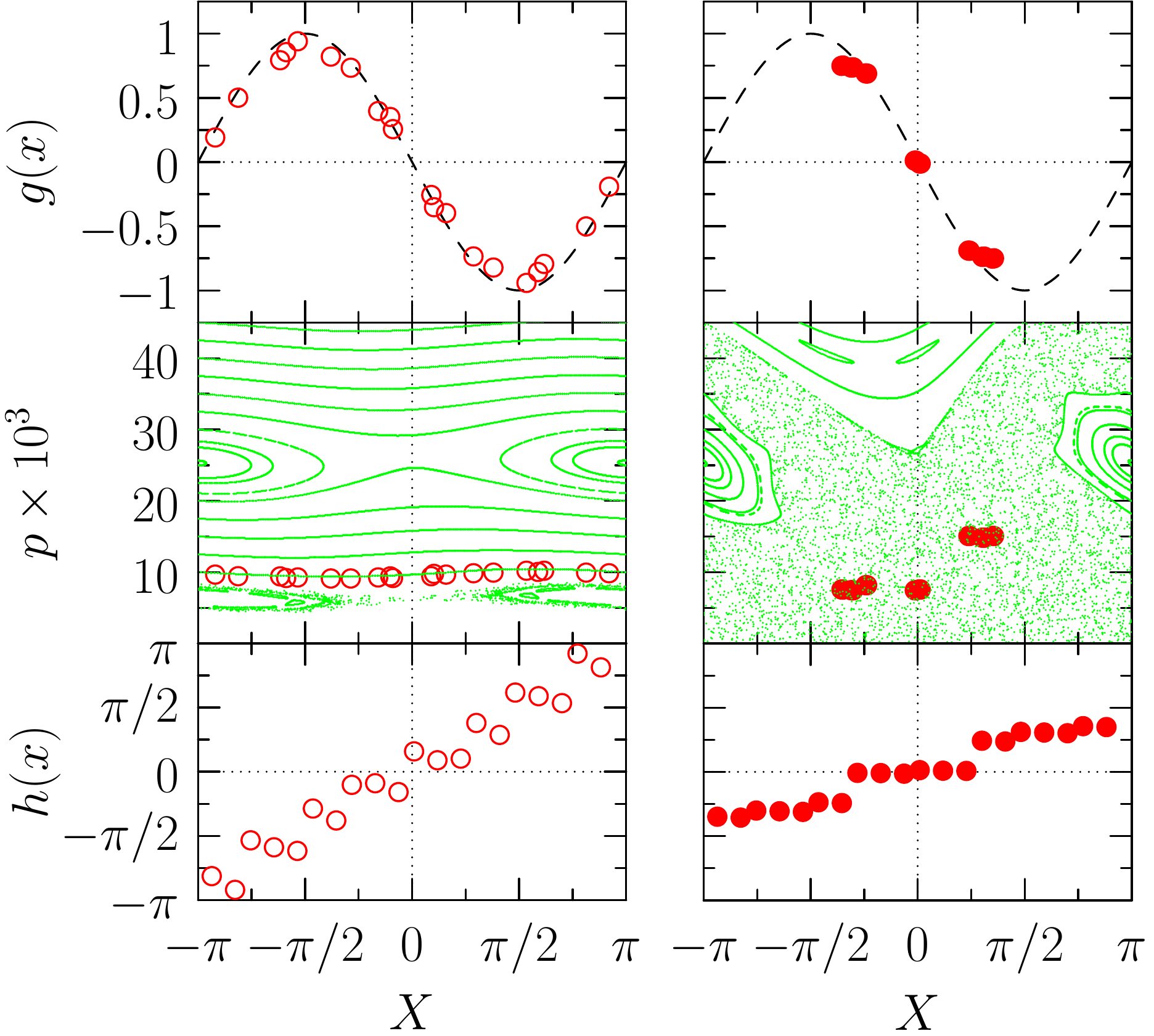}
\end{center}
\caption{\label{fig2} Functions related to the dynamical
map  obtained from the ground state equilibrium positions
$x_i$ of  ions  at $K = 0.001$ (open circles,
left column) and $K = 0.01$ (full circles, right column). 
Panels show: the kick function $g(x)$ (top); the phase space $(p, x)$
of the map  with $g(x) = - K \sin x$ (green/light-gray points) and
actual ion positions (red/gray circles) (middle); the hull function 
$h(x)$ (bottom). The ion positions are shown via
the hull function
$h(x) = (x_i+\pi) [{\rm mod }\; 2\pi]-\pi$ versus
$x=(2\pi (i-1)/\nu -\pi/2 ) [{\rm mod }\; 2\pi] -\pi$, 
for the central $1/3$ part of the chain; $i \in (N/3,\ldots,2N/3)$.
Here $N/L=60/89$.
 }
\end{figure}

As in \cite{fki2007} the validity of the map description
is  checked numerically finding 
the ground state configuration using numerical methods
of energy minimization 
described in \cite{fki2007,aubry}.
In these simulations the Coulomb interactions between all ions
are considered. Also we use the hard-wall boundary conditions
at the ends of ion chain assuming that in the experiments it can be
created by specific laser frequency detuning from
resonant transition between ion energy levels.
Due to these boundary conditions 
the local ion density is  inhomogeneous since an ion
near boundary vicinity has more pressure from other
ions in the chain (a similar inhomogeneous 
local density $\nu(x_i) =2\pi/(x_{i+1} - x_i)$
appears for ions inside a global oscillator 
potential of a trap as discussed in \cite{fki2007}).
To avoid the non-homogeneity density effect we
select the central
part of the chain with approximately $1/3$ of all ions
where the density is approximately constant.
The examples of ion density for KAM sliding and Aubry pinned
phases are shown in Fig.~\ref{fig1}. The data show that
the distribution of ions is relative smooth in the sliding phase
and is rather peaked in the pinned phase.
The selected density is chosen to be
 close to the inverse of the golden mean value
$\nu =  \tilde{\nu}_g = \nu_g^{-1}=\nu_g-1 = (\sqrt{5} -1)/2 =0.618...$
which is often  used for
 stability analysis of KAM curves
in symplectic maps (see e.g. \cite{lichtenberg,meiss}).
With the choice $N/L=55/89$, typical for common Frenkel-Kontorova model as a rational
approximant to the golden mean, we have actually $\nu=0.56$, and we need to adjust the number 
of ions $N$ in order to approach the required value $\tilde{\nu}_g=0.618...$.

We note that this  problem of local density change 
disappears if one restricts himself by the account for the nearest neighbors 
interactions between ions: in this approach the ions 
density becomes practically constant along the whole chain.
This approach is used below in our simulations of kinetic properties of the system.

The analysis of ground state stationary positions
of ions in a periodic potential is shown
in Fig.~\ref{fig2} for the KAM sliding phase ($K=0.001$)
and the Aubry pinned phase ($K=0.01$)
at $N/L=60/89$ ratio which
 gives approximately the golden mean fraction
$\nu \approx \tilde{\nu}_g \approx 0.618$ in the middle
part of the chain. Even if the Coulomb interactions 
between all ions are considered the kick function
$g(x) $ is close to the theoretical one $g(x)=-\sin x$.
This shows that the approximation of nearest neighbors
used in the symplectic map (\ref{eq:map})
gives us a good approximation of the system.
For $K=0.001$ the positions of ions follow
the invariant KAM curve while for
$K=0.01$ the ion positions form an invariant Cantor set (cantori)
appearing on the place of destroyed KAM curve.
The hull function is defined as $h(x) = (x_i+\pi) [{\rm mod }\; 2\pi]-\pi$
with $x=(2\pi (i-1)/\nu -\pi/2 ) [{\rm mod }\; 2\pi] -\pi$, where $i \in (N/3,\ldots,2N/3) $.
Indeed, at $K \rightarrow 0$
we have  $h(x)= x$ with a smooth deformation for small perturbations
while in the pinned phase we obtain the devil's staircase
corresponding to the fractal cantori structure.
The transition from the smooth hull function
to the devil's staircase is visible in the data of Fig.~\ref{fig2}
even if there is a certain spreading of points especially 
in the KAM phase which we attribute to a rather small value of $K=0.001$.
Thus the obtained data show that 
with $\nu \approx 0.618$ 
the Aubry transition takes place at a
certain $K_{c}(\nu)$ inside the interval $0.001 < K_{c}(\nu) < 0.01$.

It is important to note that a similar approximate
symplectic map description for static charge positions
is working well not only in the case of periodic potential but also for
wiggling channel structures analyzed in \cite{snake}.

\begin{figure}[t]
\begin{center}
\includegraphics[width=0.48\textwidth]{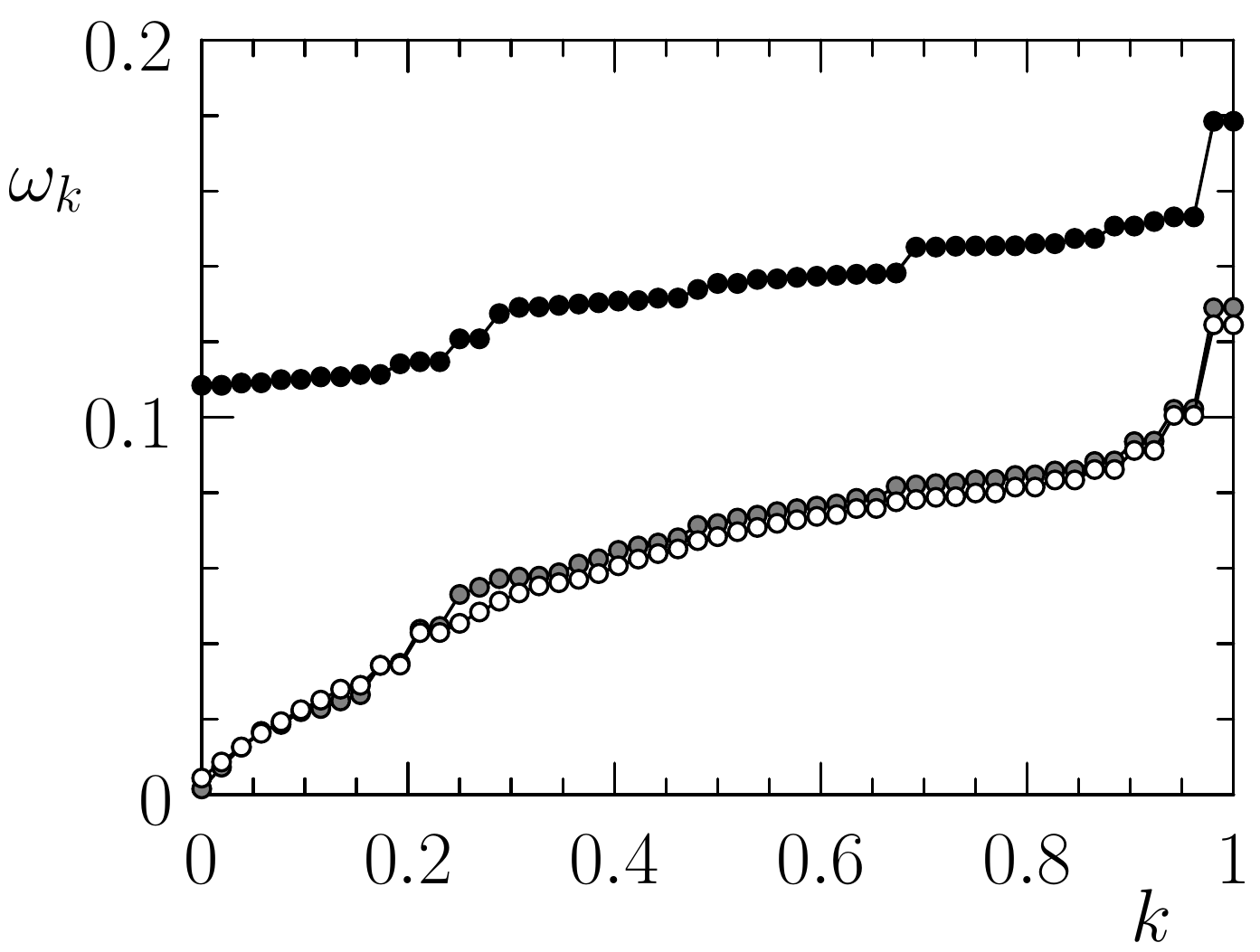}
\end{center}
\caption{\label{fig3} Spectrum of phonon excitations $\omega(k)$
as a function of scaled mode number $k=i/N$  
($i=1,\ldots,N-2 $), for
% the two ions at the chain ends are fixed !!!, therefore the number of dergees of freedom is N-2
$K=0.001$ (bottom curve, open circles), 
$K=0.002$ (bottom curve, gray circles)
 and 
$K=0.015$ (top curve, full circles). 
Here there are $N=55$ ions and $L=89$ lattice periods.
 }
\end{figure}

\begin{figure}[t]
\begin{center}
\includegraphics[width=0.48\textwidth]{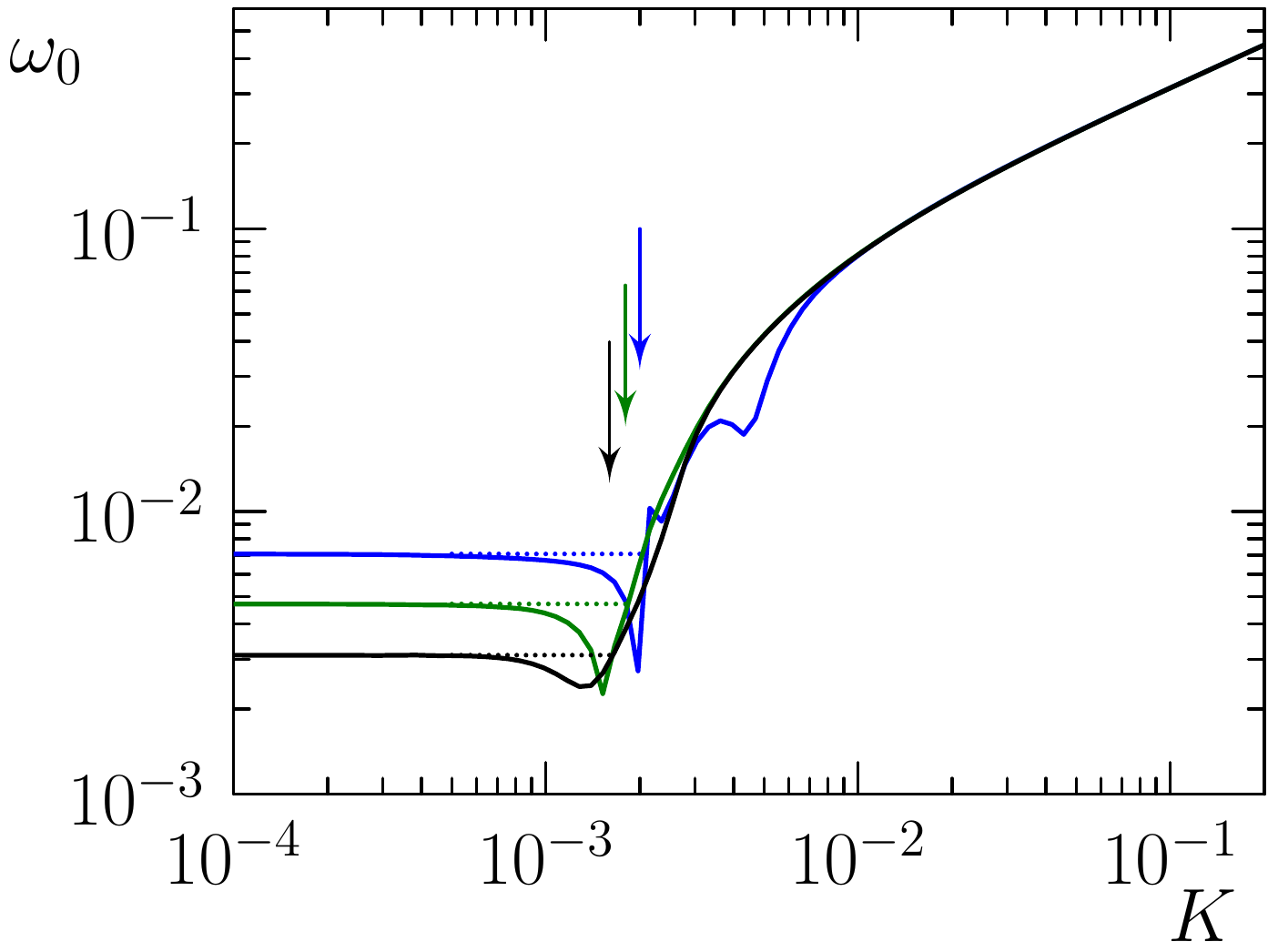}
\end{center}
\caption{\label{fig4} The lowest mode phonon frequency
$\omega_0(K)$ vs the periodic potential coupling $K$.
Blue, green and black curves correspond to 
$N/L =34/55, 55/89$ and $89/144$.
The position of the Aubry transition $K=K_c$ 
indicated by arrow
is defined as an intersection of 
the  curve $\omega_0(K)$ with the level shown by the 
dotted line at $\omega_0(K=0)$.
Three arrows correspond to different
winding numbers (densities) $ \nu = 34/55, 55/89 \text{ and } 89/144$
and correspond to $K_c=0.002, 0.0018 \text{ and }0.0016$.
 }
\end{figure}

To obtain the critical amplitude in a more exact way
we determine the static ion configuration
at a fixed density $\nu=N/L$ and then compute the phonon spectrum
$\omega(k)$
of small ion oscillations near the equilibrium positions.
Such an approach was already used and described in \cite{aubry,fki2007}. 
An example of the photon spectrum $\omega(k)$ 
is shown in Fig.~\ref{fig3} for $N/L = 55/89$. Below the transition
at $K < K_{c}(\nu)$ we have the linear acoustic type 
spectrum $\omega \approx C_{v} k$ of phonon type 
excitations describing ion oscillations
near their static equilibrium positions 
like those shown in Fig.~\ref{fig2} at $N/L=60/89$.
Here, $k=i/N$ plays the role of the wave vector and
$C_v \approx 0.2$ is the sound velocity.
The lowest phonon frequency goes to zero with the increase
of the system size as $\omega(k=1/N) \approx  C_v/N$.
The spectrum becomes drastically different
above the transition $K > K_{c}(\nu)$
with appearance of the optical gap $\omega_0$
in the spectrum $\omega(k)$.
In fact, the gap $\omega_0$ is proportional to the Lyapunov exponent
of symplectic map orbits located on the corresponding 
fractal cantori set \cite{aubry,lichtenberg}.
It is independent of the system size $N$. 

In principle, it is possible to
search numerically for the value of $K_{c}(\nu)$
at which a nonzero optical gap
appears in the spectrum $\omega(k)$.
However, we find more useful
to compute the dependence of lowest frequency
$\omega_0 = \omega(k=1/N)$ on the amplitude
$K$. In the KAM phase we have $\omega_0(K)$ being
independent of $K$ and thus we determine the critical
$K_{c}(\nu)$ by the intersection of
horizontal line $\omega_0(K) = \omega_0(K=0)$
with the curve of the spectrum $\omega_0(K)$ at 
higher $K$ values ($\omega_0(K=0) = \omega_0(K)$).
The example of this intersection procedure
is shown in Fig.~\ref{fig4}
at increasing Fibonacci approximates
$N/L=34/55, 55/89, 89/144$.
We estimate the accuracy of this method 
of $K_c$ computation on the level of 10\%-15\%
of $K_c$ value. At small sizes like $34/55$ 
a wiggling of curve at higher $K$ decreases the accuracy.
But at longer sizes the result is stable.
Thus we obtain the critical $K_{c}(\nu) \approx 0.0017$
for  the irrational density
$N/L=\nu=\nu_g  =0.618... \;$,
taking the average value for sizes $55/89$ and $89/144$.

\begin{figure}[t]
\begin{center}
\includegraphics[width=0.48\textwidth]{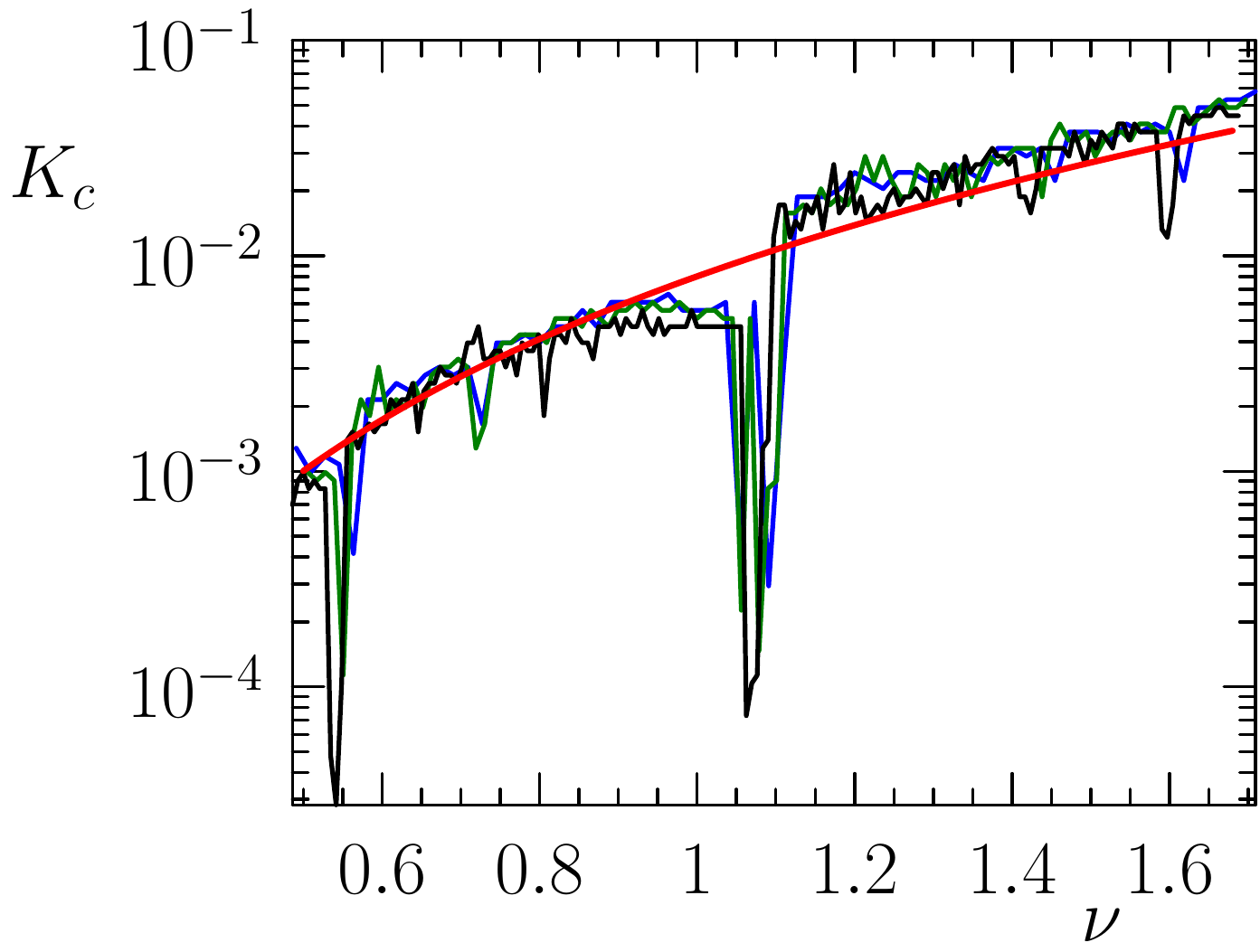}
\end{center}
\caption{\label{fig5} Dependence of 
the critical coupling $K_c$ on the winding number $\nu=N/L$.
Blue, green and black curves correspond to $L=55, 89$ and $144$;
for each $L$ the number of ions $N$ takes all integer values
up to corresponding highest/lowest $\nu$ value.
The red curve corresponds to theoretical estimate: $K_c=0.034 (\nu/1.618)^{3}$.
 }
\end{figure}

The numerically obtained dependence 
of the critical amplitude of 
Aubry transition $K_{c}(\nu)$ on 
ion density $\nu$ is shown in Fig.~\ref{fig5}
for the range $0.5 \leq  \nu \leq 1.7$.
The numerical values $K_{c}(\nu)$ are obtained as in Fig.~\ref{fig4}
for the lattice size $L=55, 89, 144$ and for all integer $N$ values
appearing inside the interval $0.5 \leq \nu \leq 1.7$.
On average the numerical values are in a good agreement
with the theoretical expression given in \cite{fki2007}
\begin{eqnarray}
K_{c}(\nu) \approx 0.034 (\nu/\nu_g)^3 \; , \;\; \nu_g=1.618... \; .
\label{eq:kc}
\end{eqnarray} 
This formula is obtained on the basis of local reduction of map
(\ref{eq:map}) to the Chirikov standard map with a homogeneous
density of nonlinear resonances. For that the second equation (\ref{eq:map})
is linearized near $1/\sqrt{p} = 2\pi/\nu_g$ 
that gives the Chirikov standard map with chaos
parameter $K_{\rm eff} = K(2\pi/\nu_g)^3/2$ and 
the critical value $K_c = 0.034$ at 
$\nu_g=1.618...$ corresponding to $K_{\rm eff} \approx 1$. 
More details of this  method are given in
\cite{fki2007,chirikov,lichtenberg,scholarchi}).
For $\nu=\nu_g=1.618...,$ the detailed numerical analysis gives 
$K_{c}(\nu)  \approx 0.0462$ \cite{fki2007,ztzs} (instead of $0.034$)
and  for $\nu=0.618...$ we have $K_{c}(\nu) \approx 0.0017$  (instead of $0.0189$).
The strongest deviations from the theoretical dependence (\ref{eq:kc})
take place at $\nu \approx 0.55$ and $\nu \approx  1.06$
with a significant drop of $K_c$. In fact, these $\nu$ values
are located in the vicinity of main resonances with $\nu=1/2; 1$
which are slightly shifted from their rational positions due to
inhomogeneity of ion density appearing due to 
the hard-wall boundary conditions
(see Fig.~\ref{fig1}).  In a vicinity of rational resonances
there is a stochastic layer inside which the 
dynamical chaos emerges at rather small critical perturbations
(see \cite{chirikov,lichtenberg}).

In the following the analysis of the thermodynamic characteristics
is done for $\nu \approx 0.618$. In the related Figs.~\ref{fig7}-\ref{fig11}
for rescaling of parameters to the dimensionless units
we use $K_c=0.002$ which is close to the theoretical value
$K=0.00189$ from (\ref{eq:kc}).

Thus in global the numerical results obtained here for 
the dependence $K_c(\nu)$ on density are in good agreement with the theory 
(\ref{eq:kc}) developed in \cite{fki2007}.
The most striking feature of this dependence is 
the sharp decrease of $K_c$ with $\nu$.
This allows to observe the Aubry transition
at significantly smaller laser powers
simply by a moderate decrease of density $\nu$ 
(see discussion below).

\begin{figure}[t]
\begin{center}
\includegraphics[width=0.48\textwidth]{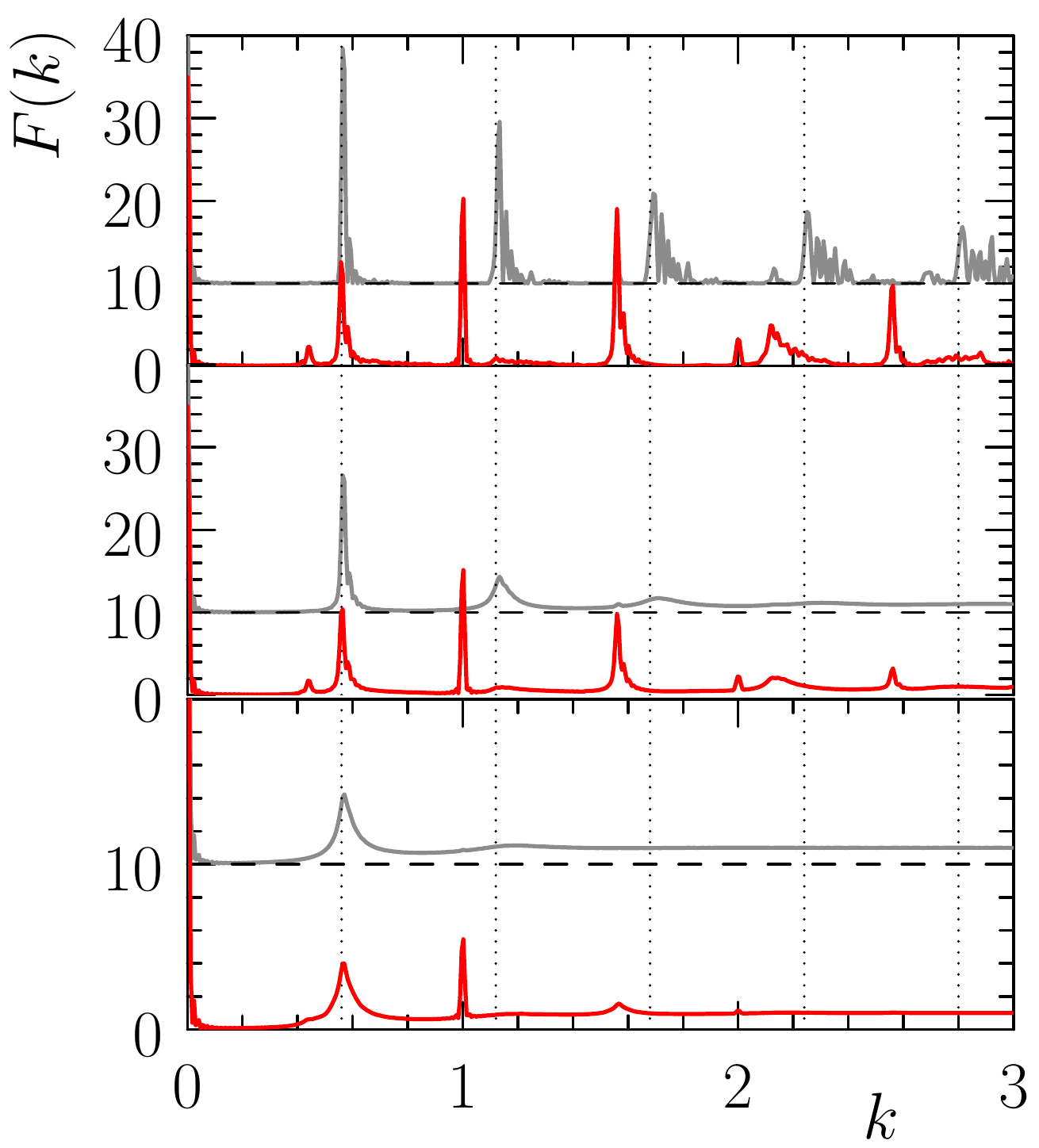}
\end{center}
\caption{\label{fig6} Dependence of the formfactor
$F(k)$ on $k$ below the Aubry transition ($K=0.001<K_c=0.00189$) 
(KAM sliding phase, gray curves)
and above ($K=0.01>K_c=0.002$) (pinned mode, red curves).
For a clarity  the gray curves are shifted up by 10.
The temperature takes values $T=10^{-6},2\cdot10^{-3}, 10^{-2}$ for 
top, middle and bottom panels respectively.
Here $N/L=55/89$.
Dotted curves indicate the peak positions at $k_i=\nu i$, where $\nu=0.56$ is
the mean rotation number at the central region of the chain.
(compare this value of $\nu$ to that in the Fig.\ref{fig1}).
The formfactor is calculated over central part of ion chain, 
contributions from 10 ions from each edge are ignored.
 }
\end{figure}

\section{Formfactor of the ion structure}
\label{sec:4}

It is useful to characterize the structure of ions in a 
periodic potential by the formfactor $F(k)$ given by
\begin{eqnarray}
 F(k) =\left\langle \left|  \sum_{i \neq j}^{N_{cr}} \exp\left(i k \left(x_i(t)-x_j(t)\right)\right) \right|^2 \right\rangle/N_{cr} \; ,
\label{eq:formfactor}
\end{eqnarray}
where the sum is taken by the central part of the chain with 
approximately $N_{cr} \approx N/3$. 
The positions of ions $x_i(t)$ at different moments of time
are obtained at finite temperature $T$ with the Langevin equations (\ref{eq:langevin})
with averaging over time. Here we use the hard-wall boundary conditions
and Coulomb interactions act between all electrons ions.
Here we use the Metropolis algorithm for simulation of the thermal
effects (see e.g. \cite{fki2007,snake}), rather than the Langevin equation,
and the nearest neighbor approximation is not necessary.

The variation of $F(k)$ 
with temperature is shown in Fig.~\ref{fig6} for
the KAM sliding phase and the Aubry pinned phase.
In the KAM phase at $K<K_c$ the peaks of $F(k)$ are located
at $k \approx i \nu$ corresponding to 
integer harmonics $i$ of average ion density $\nu \approx  0.56$
in the central part of the chain.
In contrast, in the Aubry phase at $K>K_c$  
there appear additional integer peaks at 
$k \approx i$ being
commensurable with the lattice period.
At low temperature $T = 10^{-6}$ the peaks of $F(k)$ are well visible
and clearly demonstrate the transition from
incommensurate phase at $K<K_c$
to the quasi-commensurate phase above the Aubry transition at $K>K_c$.
with the increase of temperature the high harmonics
in $F(k)$ are washed out by thermal fluctuations.
We think that the analysis of formafactor structure
can be rather useful for experimental investigations
of Aubry transition.

\begin{figure}[t]
\begin{center}
\includegraphics[width=0.48\textwidth]{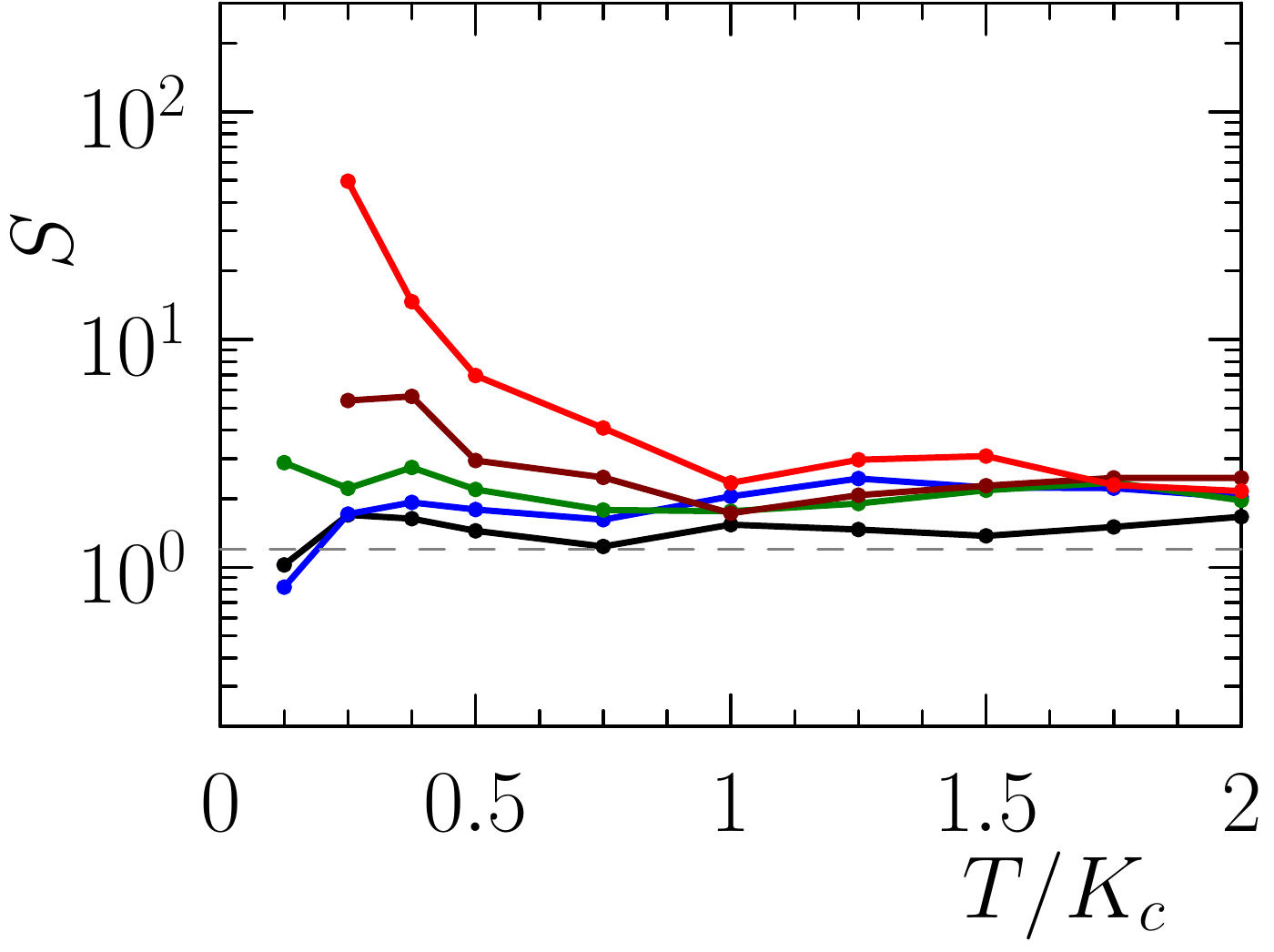}
\end{center}
\caption{\label{fig7} Dependence of Seebeck coefficient $S$ on 
temperature $T$ at different values
$K/K_c = 0$ (black), $1$ (blue), $2$ (green), $3$ (brown), $4$ (red).
Here the ratio of number of $N$ ions 
to number of lattice periods $L$ is $N/L=21/34$
and $K_c = 2\times 10^{-3}$. Dashed line shows
$S=1.2$ value at $K=0$ for noninteracting particles.
 }
\end{figure}

\section{Seebeck coefficient}
\label{sec:5}

The computations of the Seebeck coefficient $S$ are done in the frame
of Langevin equations (\ref{eq:langevin})
with the hard-wall boundary conditions.
We note that in the Aubry pinned phase 
long computational times are 
required. Thus we made simulations with times up to $t=10^8$.
In these simulations for each ion we take into account 
the Coulomb interactions only with nearest left and right neighbor ions
that allow to accelerate the computations.
As in \cite{ztzs} we ensured that the obtained results are not sensitive
to inclusion of interactions with other neighbors.

\begin{figure}[t]
\begin{center}
\includegraphics[width=0.48\textwidth]{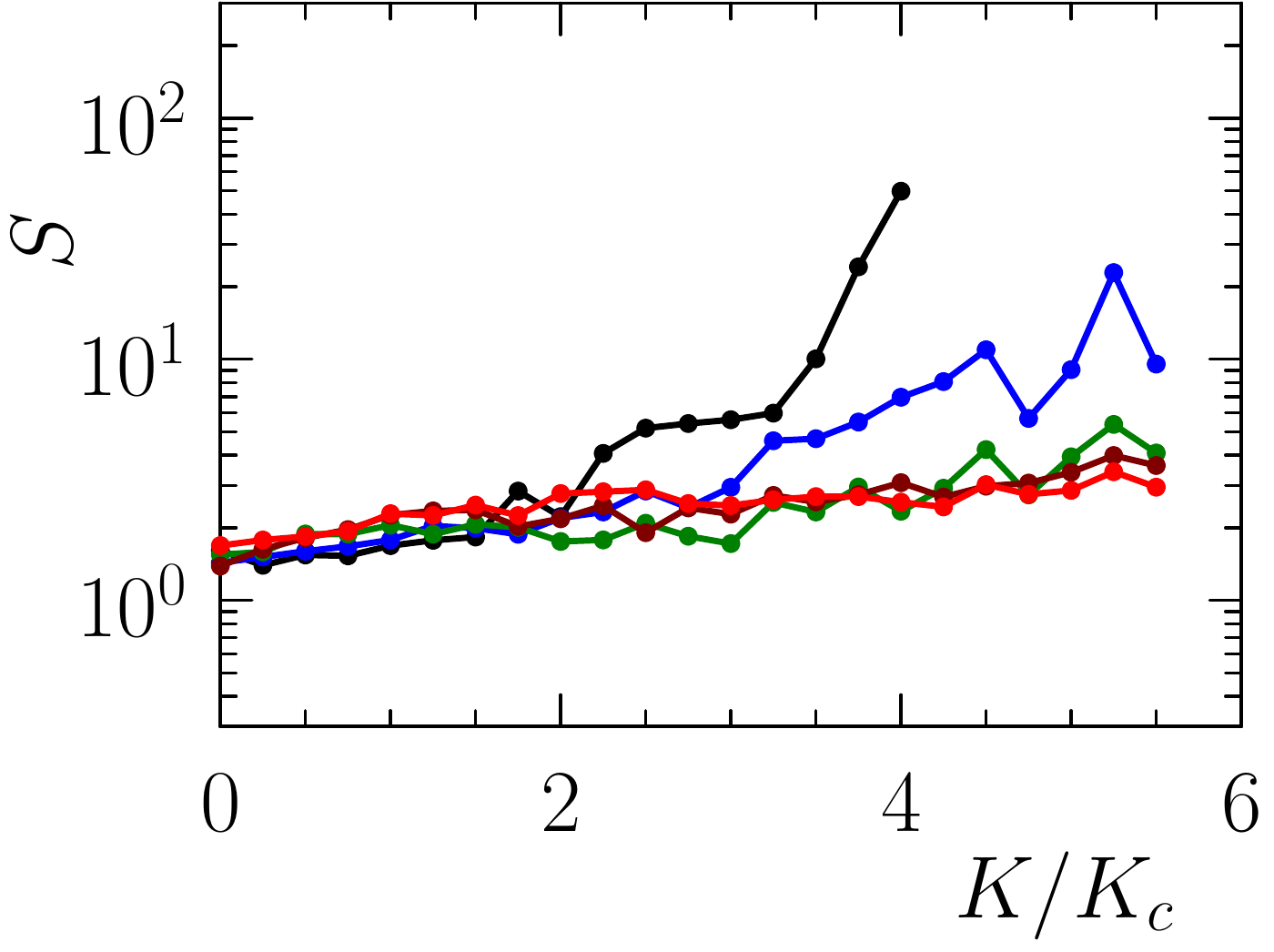}
\end{center}
\caption{\label{fig8} Seebeck dependence on
$K$ at different temperatures 
$T/K_c=0.25$ (black),
$0.5$ (blue), $1$ (green), $ 1.5$ (cyan),
$2$ (red). 
Here the ratio of number of $N$ ions 
to number of lattice periods $L$ is $N/L=21/34$
and $K_c = 2 \times 10^{-3}$. 
 }
\end{figure}

The computations of $S$ are done with the procedure
developed in \cite{ztzs}. At fixed temperature $T$
we apply a static field $E_{dc}$ which
creates a voltage drop 
$\Delta V = E_{dc} 2\pi L$ and
a gradient of ion density $\nu(x)$
along the chain. Then at $E_{dc}=0$ within
the Langevin equations (\ref{eq:langevin}) 
we impose a linear gradient of temperature
$\Delta T$  along the chain and in the stabilized 
steady-state regime determine the charge density
gradient of $\nu(x)$ along the chain
(see e.g. Fig.2 in \cite{ztzs}).
The data are obtained in the linear regime
of relatively small $E_{dc}$ and $\Delta T$ values.
Then the Seebeck coefficient
is computed as $S=\Delta V/\Delta T$
where $\Delta V$ and $\Delta T$
are taken  at such values
that the density gradient from $\Delta V$  
compensates the one from $\Delta T$.

The dependencies of $S$ on temperature $T$ at
different amplitudes $K$ of the periodic potential
are shown in Fig.~\ref{fig7}. In the KAM sliding
phase with $K \leq K_c$ there is no significant change
of $S \approx 2$, which remains close to its value
$S=1.2$ for free noninteracting particles.
In contrast in the Aubry pinned phase with $K/K_c > 1$
we obtain an exponential increase 
of $S$ at $T/K_c < 0.5$. 
This increase of $S$  is especially
visible for $K/K_c=4$ where the maximal
reached value is $S \approx  50$.

We note that a similar strong growth of $S$
at low temperatures has been reported
in experiments with quasi-one-dimensional conductor
$\rm(TMTSF)_2PF_6$ \cite{espci} where 
as high as $S =400 $ value had been reached at low temperatures
(see Fig.~3 in \cite{espci}).

The dependence of $S$ on $K$ is shown in 
Fig.~\ref{fig8}. At low temperatures $T<K_c$
there is a sharp increase of $S$ with increase of $K>K_c$
with a maximal reached value $S \approx 60$.
In a certain sense the increase
of $K$ drives the system deeper and deeper
into the insulating phase with larger and larger 
resistivity. Thus we can say that $S$ increases
with the increase of sample resistivity.
A similar dependence has been observed 
in experiments with two-dimensional
electron gas in highly resistive
(pinned) samples of micron size 
(see Fig.~8 in \cite{pepper}).

Thus we can say that our numerical results are in a qualitative
agreement with the experiments \cite{espci,pepper}.

\section{Figure of merit $ZT$}
\label{sec:6}

\begin{figure}[t]
\begin{center}
\includegraphics[width=0.48\textwidth]{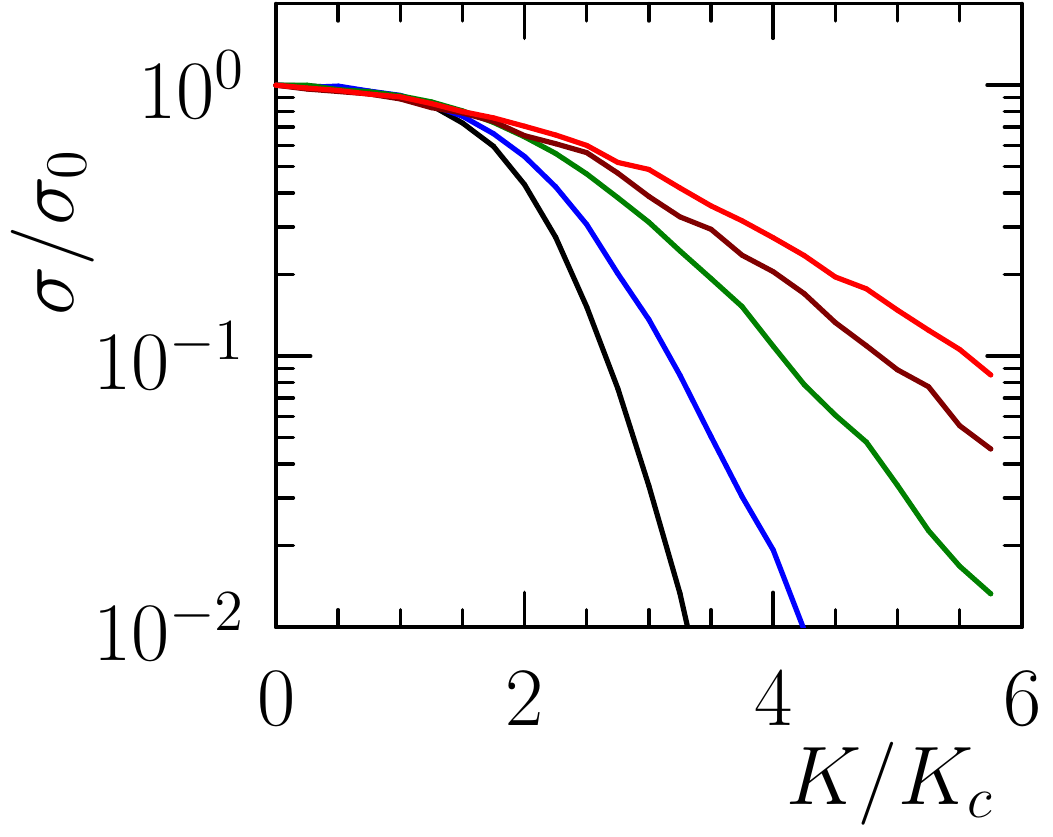}
\end{center}
\caption{\label{fig9} Dependence of conductivity on 
$K$ at different temperatures 
$T/K_c=0.25$ (black),
$0.5$ (blue), $1$ (green), $ 1.5$ (brown),
$2$ (red). 
Here the ratio of number of $N$ ions 
to number of lattice periods $L$ is $N/L=21/34$,
$\eta=0.02$, $\sigma_0 =\nu/(2\pi\eta)$ with
$\nu=0.618$
and $K_c = 2 \times 10^{-3}$.
 }
\end{figure}

To determine the figure of merit $ZT$ we need to compute in addition to $S$
the electrical conductivity $\sigma$ 
and the heat conductivity $\kappa$. 
The computation of $\sigma$ is done
by applying a weak $E_{dc}$ field acting on a circle
with periodic boundary conditions.
Then we compute the ion velocity $v_{ion}$
averaged over all particles and time interval.
Then the charge current is $j=\nu v_{ion}/2\pi$ 
and $\sigma=j/E_{dc}$. At $K=0$ the ion crystal moves
as a whole with $v_{ion}=E_{dc}/\eta$ corresponding to the conductivity
of free particles $\sigma=\sigma_0=\nu/(2\pi \eta)$
that is confirmed by numerical simulations
(see Fig.~\ref{fig9}).

The dependence of $\sigma$ on $K$ and temperature $T$
is shown in Fig.~\ref{fig9}. In the KAM sliding phase at $K<K_c$,
$\sigma/\sigma_0$ is practically independent of 
$T$ and $K$. However, in the Aubry pinned phase at $K> K_c$
there is an exponential drop of $\sigma/\sigma_0$
with increase of $K$ and with decrease of $T$.

For the computation of heat conductivity
$\kappa$ we use the approach developed in \cite{ztzs}.
The heat flow $J$ is related
to the temperature gradient 
by the Fourier law with the thermal
conductivity $\kappa$: $J=\kappa \partial T / \partial x$
\cite{thermobook,politi}. The heat flow is computed from 
forces acting on a given ion $i$ 
from left and right sides being respectively
$f_i^{L}=\sum_{j<i} 1/|x_i-x_j|^2$,
$f_i^{R}=-\sum_{j>i} 1/|x_i-x_j|^2$.
For an ion moving with a velocity $v_i$
these forces create  left and right energy flows 
$J_{L,R} = \langle f_i^{L,R} v_i \rangle_t \;$.
In a steady state the mean ion energy is independent of 
time and $J_L + J_R=0$. But the difference of these flows
gives the heat flow along the chain: 
$J=(J_R-J_L)/2 = \langle ( f_i^{R}- f_i^{L}) v_i/2 \rangle_t \; $.
We perform such  computations of the heat flow,
with hard wall boundary conditions.
In addition to the method used in \cite{ztzs}
we perform time averaging
{using accurate numerical integration along the ions trajectories
that cancels contribution of large oscillations due to periodical motion of ions.}
In this way we  determine the thermal conductivity
via the relation $\kappa=J L/\Delta T$.
The obtained results for $\kappa$ are independent
of small $\Delta T$. Our previous result 
confirm that the thermodynamic characteristic 
are independent of system size \cite{ztzs}.

\begin{figure}[t]
\begin{center}
\includegraphics[width=0.48\textwidth]{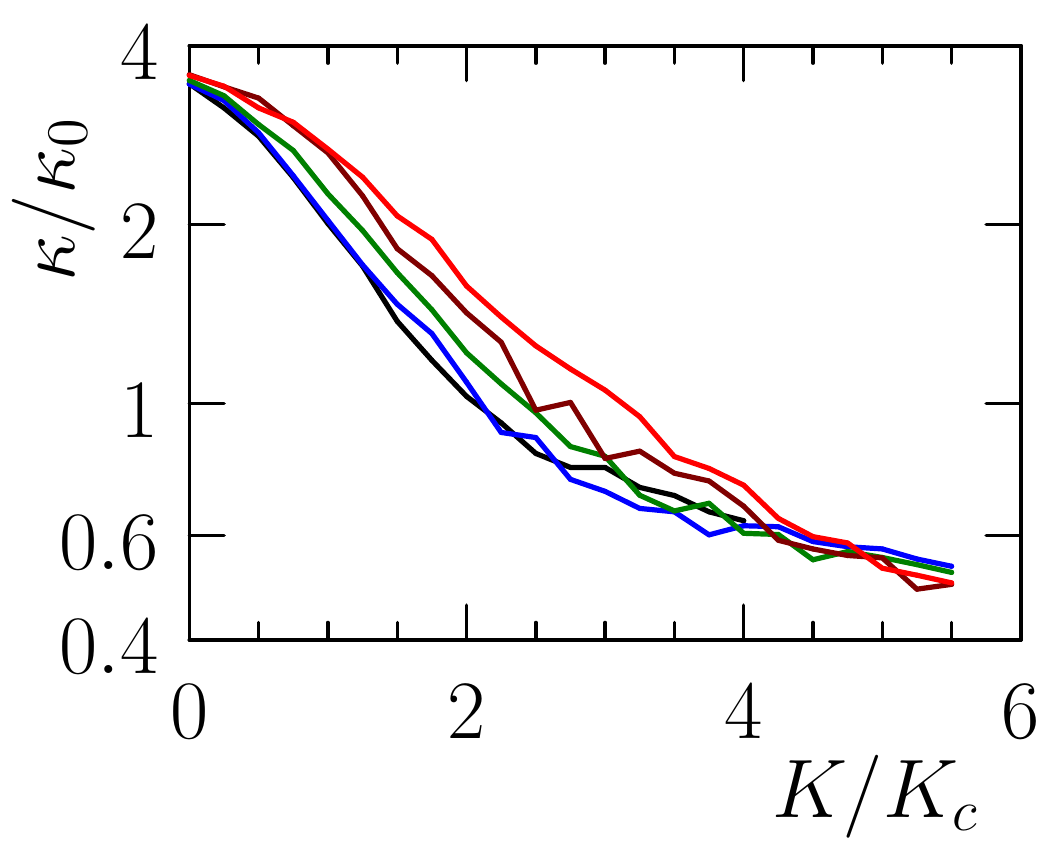}
\end{center}
\caption{\label{fig10} Dependence of heat conductivity $\kappa$ on 
$K$ at different temperatures 
$T/K_c=0.25$ (black),
$0.5$ (blue), $1$ (green), $ 1.5$ (brown),
$2$ (red). 
Here the ratio of number of $N$ ions 
to number of lattice periods $L$ is $N/L=21/34$,
$\eta=0.02$, $\kappa_0=\sigma_0 K_c$
and $K_c = 2 \times 10^{-3}$.
 }
\end{figure}

The dependence of heat conductivity $\kappa$ 
on $K$ and $T$ is shown in Fig.~\ref{fig10}.
For convenience we present the ratio of $\kappa$ to 
$\kappa_0=\sigma_0 K_c$ to have the results in dimensionless units.
There is an exponential decrease of $\kappa/\kappa_0$ with increase of $K>K_c$
showing that for ionic phonons the propagation along the chain becomes
more difficult at high $K$ amplitudes. The decrease of $T$ 
leads to a decrease of $\kappa/\kappa_0$ but this drop is 
less pronounced comparing to those for $\sigma/\sigma_0$.

\begin{figure}[t]
\begin{center}
\includegraphics[width=0.48\textwidth]{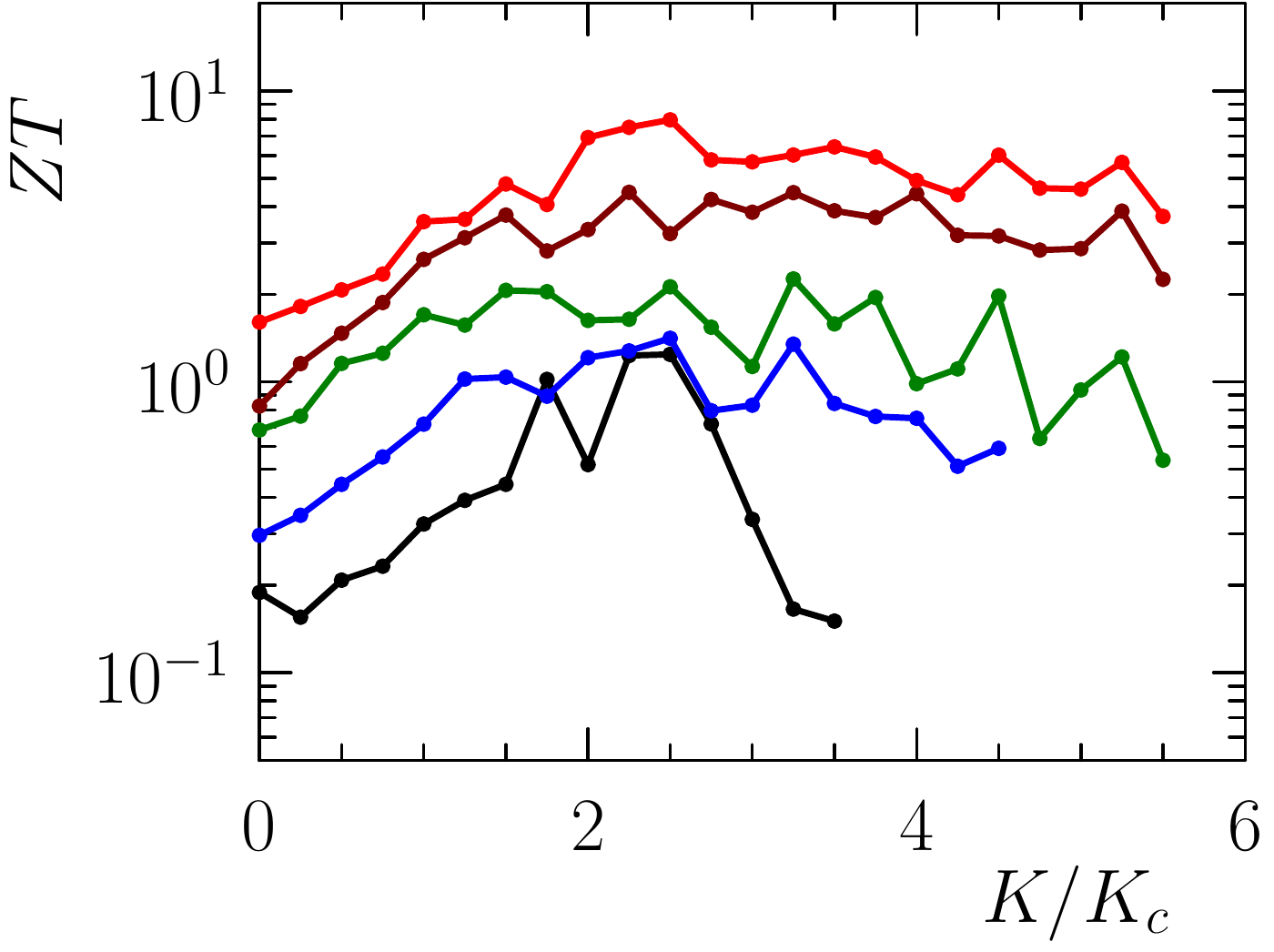}
\end{center}
\caption{\label{fig11} Dependence of figure of merit
$ZT$ on 
$K$ at different temperatures 
$T/K_c=0.25$ (black),
$0.5$ (blue), $1$ (green), $ 1.5$ (brown),
$2$ (red). 
Here the ratio of number of $N$ ions 
to number of lattice periods $L$ is $N/L=21/34$, $\eta=0.02$
and $K_c = 2 \times 10^{-3}$.
 }
\end{figure}

From the obtained values of $S$, $\sigma$ and $\kappa$ at $\nu \approx 0.618$
we compute the figure of merit $ZT$ shown in Fig.~\ref{fig11}
as a function of $K/K_c$ at different temperatures.
At fixed $T$ there is an optimal maximal $ZT$ value
located approximately at $K/K_c \approx 2.5$
while at smaller and higher $K/K_c$ values 
we have a decrease of $ZT$.
The maximal $ZT$ value increases with the growth of temperature
and the width of the peak becomes broader.
At the maximum with $T/K_c=2$ and $K/K_c=2.5$ we obtain
$ZT \approx 8$. This maximal $ZT$ value is by factor $2$ larger
than the value obtained at $\nu \approx 1.618$, $T/K_c \approx 3$ and 
$K/K_c \approx 3.5$ in \cite{ztzs}.
We suppose that a more dilute ion density
favors more efficient thermodynamical characteristics.

Our studies are done for the strictly 1D case. It is important to consider
extensions to the two-dimensional (2D) case which may be useful for electrons on liquid
helium and material science. Indeed, in quasi-1D case it is known 
that transfer dynamics can lead to excitation of transfer modes and zigzag instabilities
\cite{morigi,peeters}. However, the recent results reported for 2D
Wigner crystal transport indicate that there is similarity between 1D and 2D properties 
and thus we expect that the obtained results will be useful for 
higher-dimensional systems.

\section{Discussion}
\label{sec:7}

In this work we analyzed the properties of 
cold ion chain placed in a periodic potential.
The results show the emergence of Aubry
transition from KAM sliding phase
to Aubry pinned phase when the amplitude of the potential exceeds
a critical value $V_{A}=K_c(\nu)e^2/(\ell/2\pi)$.
For a typical period of optical lattice
$\ell=1\rm\mu m$ and dimensionless ion density per period
$\nu = 1.618$ this corresponds to
$V_A \approx 3 \rm K$ (Kelvin) that would 
require rather strong laser power for creation
of such optical potential.
However, with a decrease of $\nu$ we have a cubic drop
of $V_A$ and thus for $\nu =0.38$ we need
only $V_A \approx 0.04 \rm K$
that is much more accessible for optical lattice experiments.

The interest of the Aubry phase is related to high
thermodynamic properties appearing in this phase. Thus 
we show that one can reach in this 
phase the  high values of Seebeck coefficient 
$S \approx 50 k_B/e \approx  4400 \rm\mu V/K$
which approaches to the record experimental values $S \approx 400 k_B/e$
observed in quasi-one-dimensional materials \cite{espci}
and with two-dimensional electron gas in small disordered samples 
with $S \sim 50 k_B/e$ \cite{pepper}.
Even more remarkable is that in this phase 
the figure of merit can be as such high as
$ZT \approx 8$ being above the record experimental values \cite{ztsci2017}.
We suppose that for cold ions in optical lattices
the voltage difference $\Delta V$ can be easily created
by a weak external static electric field
while the temperature difference $\Delta T$
at the ends of the lattice
can be generated by additional laser heating.
Thus such experiments would allow to investigate the
thermodynamical properties of Wigner crystal
of cold ions in optical lattice. 
We assume that the investigations
of thermoelectric properties with cold ions in optical lattices
may bring us to a deep understanding of
thermoelectricity which will be further used 
for selection of optimal thermoelectric materials.

Also we think that the simple models considered here
rise an important challenge for computational methods of
quantum chemistry where interactions of electrons and atoms
are taking into account in the computations of band 
structures but after that the thermoelectric characteristics
are computed for effectively noninteracting electrons 
(see e.g. \cite{kozinsky,melendez}). We argue
that our results clearly show that the high
thermoelectric characteristics appear only in the Aubry pinned phase
where the interactions between charges play a crucial role.
We argue that our results  challenge the further development of methods
of quantum chemistry.

Thus the above analysis clearly demonstrates
the importance of Aubry pinned phase for
high thermoelectric performance.
Due to that the experimental investigations of the Aubry
transition with cold ions in optical lattices
and electrons on liquid helium
will bring to us important fundamental 
results useful for development of
efficient thermoelectric materials.

% Acknowledgments before appendices 
%-------------------------------
\section{ Acknowledgments}
This work was supported in part by the Pogramme Investissements
d'Avenir ANR-11-IDEX-0002-02, reference ANR-10-LABX-0037-NEXT 
(project THETRACOM);
it was granted access to the HPC resources of 
CALMIP (Toulouse) under the allocation 2017-P0110.
This work was also supported in part by the Programme Investissements
d’Avenir ANR-15-IDEX-0003, ISITE-BFC (project GNETWORKS).
For OVZ this work is partially supported by the Ministry of
Education and Science of the Russian Federation.
One of us (DLS) thanks R.V.Belosludov and V.R.Belosludov
for discussions of methods of quantum chemistry
used for computations of thermoelectric characteristics
in real materials.

\end{document}